\documentclass[reprint,amsmath,amssymb,aps,pra]{revtex4-1}
\usepackage[utf8]{inputenc}
\usepackage{mathrsfs,amsmath}
\usepackage{bm}
\def\ph2{{\it p}-H$_2$}
\def\Am3{\AA$^{-3}$}
\usepackage{ulem}
\usepackage{amsmath}
\usepackage{amsfonts}
\usepackage{graphicx}
\usepackage{appendix}
\usepackage{dsfont}
\usepackage{amssymb}
\usepackage{bm}
\usepackage{color}
\usepackage{natbib}
\usepackage{nccmath}
\usepackage{array}
\usepackage{graphicx}% Include figure files
\usepackage{dcolumn}% Align table columns on decimal point
\usepackage{bm}% bold math
\usepackage{float}
\usepackage[bookmarks=true,colorlinks,linkcolor=blue,urlcolor=blue,citecolor=blue]{hyperref}
\usepackage[mathlines]{lineno}% Enable numbering of text and display math
\usepackage[english]{babel}
\usepackage{textcomp}

\newcommand{\beq}{\begin{equation}}
\newcommand{\eneq}{\end{equation}}
\newcommand{\be}{\begin{equation}}
\newcommand{\ee}{\end{equation}}
\newcommand{\bea}{\begin{eqnarray}}
\newcommand{\eea}{\end{eqnarray}}

\renewcommand{\cite}[1]{[\onlinecite{#1}]}
\definecolor{burntorange}{rgb}{0.8, 0.33, 0.0}
\definecolor{brightpink}{rgb}{1.0, 0.0, 0.5}
\definecolor{bittersweet}{rgb}{1.0, 0.44, 0.37}

\usepackage{multirow}
\usepackage{wasysym}



\begin{document}
\title{Quasi-one-dimensional $^4$He in nanopores}
\author{Andrea Nava$^{(1,2)}$, Domenico Giuliano$^{(1,2)}$, Phong H. Nguyen$^{(3)}$,  and Massimo Boninsegni$^{(1,3)}$ }
\affiliation{
$^{(1)}$ Dipartimento di Fisica, Universit\`a della Calabria Arcavacata di 
Rende I-87036, Cosenza, Italy \\
$^{(2)}$ I.N.F.N., Gruppo collegato di Cosenza, Arcavacata di Rende I-87036, Cosenza, Italy \\
$^{(3)}$Department of Physics, University of Alberta, Edmonton, T6G 2E1, Alberta, Canada}
 
\date{\today}

\begin{abstract}
Low temperature structural and superfluid properties of $^4$He confined in cylindrical nanopores are theoretically investigated by means of first principle Quantum Monte Carlo (QMC) simulations. We vary the density of $^4$He inside the pore, as well as the pore diameter and the potential describing the interaction of each $^4$He atom with the pore surface. Accordingly, the $^4$He fluid inside the pore forms either a single channel along the axis, or a series of concentric cylindrical shells, with varying degrees of shell overlap. In the limit  of pore length greatly exceeding its radius, the $^4$He fluid {always}  displays markedly one-dimensional behavior, with no ``dimensional crossover'' above some specific pore radius and/or as multiple concentric shells form, {in contrast to what recently claimed by other authors [Phys. Rev. B {\bf 101}, 104505 (2020)]}.
Indeed, the predicted robustness of one-dimensional physics suggests that this system  may offer a broadly viable pathway to the experimental observation of  exotic behavior of, e.g., junctions of interacting Tomonaga-Luttinger liquids, in an appropriately designed network of nanopores.

\end{abstract}

\maketitle

\section{Introduction}
\label{intro}

An interacting Bose system cooled toward absolute zero is believed either to crystallize, or, if it escapes solidification, to undergo Bose-Einstein condensation (BEC), with the concomitant onset of  superfluidity (SF), both SF and BEC occurring at the same temperature $T_c$, in $d=3$ physical  dimensions
\cite{Leggett2006,Kora2020b}.

A dimensional reduction, effectively achievable experimentally in a number of ways, brings about important  modifications of this behavior.
Indeed, while in
$d=3$, at temperatures $T \le T_c$,
the one-body density matrix 
plateaus to a finite value $n_\circ$  (known as the condensate fraction) at large interparticle separations, 
in $d=2$  
it decays algebraically, while the superfluid density
shows a ``universal jump'' at $T=T_c$
\cite{Nelson1977}, in what is generically described as a Berezinskii-Kosterlitz-Thouless (BKT) 
phase transition \cite{Jose1977}. 

The situation is completely different in $d=1$ dimension, in which 
Galilean invariant (continuous)
systems cannot order, even at zero temperature. The low-energy,
 long-wavelength 
 dynamics is  
 described by the ``universal'' harmonic Tomonaga-Luttinger liquid (TLL) 
 Haldane's model (HM) \cite{Haldane1981}, 
 making stringent predictions on the behavior of several 
 observable quantities. For instance, while
 there is no superfluid phase in the thermodynamic
  ($L \to \infty$, $L$ being the system length) limit,
 superfluidity 
 manifests itself as a finite-size effect.
 According to the TLL theory, 
 the superfluid fraction $\rho_S (L,T)$ is 
 a universal function 
 of $[({L T})/v_J]$, $v_J$ being the superfluid velocity \footnote{we set the Boltzmann constant $k_B=1$}, with ${ \rho_S ( L , 0)}$
= 1 \cite{Delmaestro2010}. Furthermore, the one-body density matrix displays a power-law decay modulated by oscillations 
reflecting
the  atomic nature of the fluid at the microscopic scale \cite{Cazalilla2004,Giamarchi2004}.  Striking fingerprints of the
unique behavior are also present in the pair correlation function (PCF), which,
according to Haldane's theory,
features a power-law decay 
on top of a uniform contribution $\propto n_0^2$, where $n_0$ is the average one-dimensional (1D) particle density, and with 
higher-order harmonics of  frequencies that are multiple integers of $2 \pi n_0$. 
Correspondingly, the static structure factor  $S(k)$ exhibits a universal, linear dependence on $k$ as $k \to 0$, with a slope depending on (non-universal) TLL parameters, and peaks at wave vectors $K_l = 2 \pi l n_0$, with integer $l$ 
\cite{Haldane1981,Giamarchi2004,Citro2007,Citro2008}. 
\\ \indent
$^4$He is a paradigmatic   system displaying all the main features of SF and BEC  in Bose systems.
Thus, a $^4$He fluid confined in (quasi) 1D structures appears as a natural playground allowing for the experimental detection of the most peculiar aspects of 1D TLL physics, most of which remains yet unobserved.  
To this aim, 
several experimental avenues  have been considered to achieve the quasi-1D confinement of $^4$He, chiefly by adsorbing helium gas inside elongated cavities of nanometer size diameter, such as those that exist in a variety of porous glasses 
\cite{Sokol1996,Dimeo1998,Plantevin2001,Anderson2002,Toda2007,Prisk2013}, or nanoholes in Si$_3$N$_4$ membranes \cite{Savard2011}, as well as
carbon nanostructures  \cite{Teizer1999,Ohba2016}. 
Despite the obvious imperfections of these confining agents (defects, tortuosity, surface corrugation, etc.), the basic physics of an imbibed fluid should at least broadly mimic that taking place inside (possibly interconnected \cite{boninsegni2007}) long, smooth cylindrical channels; and, if the length $L$ of a typical channel greatly exceeds the characteristic size $R$ of confinement in the transverse direction (i.e., the  radius of the cylinder), quasi-1D behavior ought to arise.
\\ \indent
On the theory side, the phase diagram of $^4$He
confined in long cylindrical channels of radius $R$ of the order of 1 nm, has been extensively studied with 
the combined use of analytical and numerical methods \cite{Cole2000,Boninsegni2001,Delmaestro2010,Hernandez2011}. 
The behavior of the confined $^4$He fluid can be qualitatively understood taking into account basic features of interaction between two helium atoms, 
as well as of that of the atoms with the walls of the cylinder (i.e., the adsorption potential).
At sufficiently small $R$ (typically $\lesssim$ 4 \AA, although this varies depending on the adsorption potential) and realistic values of the pressure, $^4$He atoms line up on a single file along the axis of the cylinder,  zero-point excursions off the axis mainly resulting into a softening of the repulsion at short distances of the helium pair potential \cite{Omiyinka2016}. Atomic exchanges are suppressed, and the physics of the system quantitatively approaches the 1D paradigm \cite{Delmaestro2010_b,Markic2015,Markic2018}.
\\ \indent
As $R$ is increased, the structure of the imbibed fluid  for different pore fillings quite generally consists of one (or more) cylindrical (concentric) shell(s), coaxial with the pore, whose location and sharpness depend on the pressure and on the adsorption potential, upon which depends also the presence of a well-defined file of atoms along the axis; the radial distance between contiguous shells is roughly set by the repulsive core of the helium interatomic potential, of the order of 2.5 \AA\   \cite{Urban2006,Delmaestro2010,Delmaestro2010_b,Delmaestro2012,Markic2018}.
\\ \indent
Bulk three-dimensional (3D) physics must obviously emerge as both $R, L$ become macroscopic, i.e., formally in the $R, L \to\infty$ limit, with $R/L$ held constant (albeit possibly $<< 1)$  \cite{Delmaestro2010_b}. Under what conditions (quasi)-1D behavior may be observable in this geometry, especially if $R$ is large enough to allow for a multishell fluid structure inside the pore,
is a long debated issue. Based on an analysis of the axial PCF computed by QMC, Del Maestro et al. \cite{Delmaestro2010_b} concluded  
that, even for $R$ large enough to accommodate 
a few coaxial cylindrical shells in the pore, the helium fluid that fills the inner cylinder displays 1D behavior in the $L\to\infty$ limit. 
This can be understood by regarding the multishell structure as a
set of coupled bosonic TLLs
(see, for instance, Ref. \onlinecite{Orignac1998}); despite the low-energy mode gapping due to the coupling between the TLLs, the  ``center of mass mode'' survives as a gapless, low-lying degree of freedom, a signature of which is present in the 
1D behavior of the PCF (an extension of this idea, accounting for  effects of disorder, has been recently developed in Ref. \onlinecite{Yang2020}).
\\ \indent
This scenario has been recently challenged in Ref. \onlinecite{Markic2015}, where, 
based on an analysis of the scaling properties of  $\rho_S ( L , T )$, computed by QMC,
it is contended that a single cylindrical shell of $^4$He, of effective radius as tiny as 1.75 \AA, displays 2D, rather than 1D behavior. 
Specifically, their claim is that the values of $\rho_S(L,T)$
do not conform with the predictions of the TLL theory, but are  consistent instead with 2D BKT scaling. Furthermore, they argue that even in the $(R/L) \to 0$ limit there exists nonetheless a characteristic value of $R$ (presumably dependent on $n_0$ and on the adsorption potential) where a crossover takes place between 2D and 3D, i.e., the $^4$He fluid filling the pore features the same qualities of bulk $^4$He. If these conclusions were confirmed, they might conceivably deal a serious setback to the ongoing experimental effort aimed at observing signatures of TLL physics in fluids adsorbed in nanopores, as they may place daunting requirements on the confining agents, in terms of smallness of diameter size to achieve and/or type of material to utilize.
\\ \indent
In this paper, we weigh in on the above controversy and attempt to settle the issue  of the effective dimensionality of a $^4$He fluid in the confines of a cylindrical nanopore, modeled as a smooth cylindrical channel of length $L$ and   radius $R$, by performing extensive, first principle computer (QMC) simulations of a realistic model of $^4$He in cylindrical confinement. We consider pores of radius ranging from 3 to 10 \AA, as well as different adsorption potentials, allowing us to explore a variety of distinct physical settings. Specifically, we go from a single, tightly confined axial file, or a single cylindrical shell of $^4$He atoms, for pores of small radii, to systems enclosed in wider pores, in which  multiple shells form, in some cases sharply defined, with little or no particle exchanges taking places between them, in other cases lose shells, only identifiable as local maxima of the radial $^4$He density, with $^4$He atoms essentially delocalized throughout the system.
\\ \indent 
We compute the superfluid fraction $\rho_S (L,T)$ of $^4$He inside the pore, just like in Ref. \onlinecite{Markic2020}, as well as the axial static structure factor $S(k)$. As mentioned above, $S(k)$  has a known analytical form within the TLL, and is expected to display well-defined scaling properties with respect to the product $LT$ of pore length $L$ and temperature. This information can be used  to assess the degree to which our simulated systems conform to the 1D paradigm, or whether there are significant deviations in some cases.
\\ \indent
The main conclusion of our study is that, in the limit $(R/L)\to\ 0$, the 1D behavior of $^4$He is {\it   always} recovered, regardless of pore radius and/or shell structure inside the pore, in contrast with the claim of Ref. \onlinecite{Markic2020}. We see no evidence of any dimensional crossover to 2D or 3D.
We attribute the disagreement between our conclusions and those of Ref. \onlinecite{Markic2020}, to the fact that the length $L$ of the systems studied therein (making use of the same computational technology adopted here), is insufficient to reach the TLL regime \footnote{This conclusion appears to be corroborated by a comment made in Ref. \onlinecite{Markic2020}, to the effect that the agreement between their numerical data and the TTL predictions improves with increased system size, i.e., 1D behavior is approached asymptotically.}. We generally confirm both the findings of Refs. \onlinecite{Delmaestro2010_b,Delmaestro2012}, as well as the validity of the basic assumption of the theory expounded in Ref. \onlinecite{Yang2020}, namely that a single cylindrical shell is a 1D object as $L\to\infty$, lending support to the notion that the properties of a fluid in cylindrical confinement may be predicted on the basis on a formalism of coupled concentric shells. 
\\ \indent
The paper is organized as follows: in Section \ref{nparham}, we briefly discuss the microscopic $N$-particle Hamiltonian for $^4$He in a nanopore; 
in Section \ref{4hepores} we provide and discuss our results, while in Section \ref{conclusions} we outline 
 our conclusions and provide some further future perspectives of our work.
\section{Model and Methodology}
\label{nparham}
We model the system of interest as an ensemble of $N$ $^4$He atoms, regarded as point particles of mass $m$ and spin zero, i.e., obeying Bose statistics. The microscopic many-body Hamiltonian reads as follows:

\beq
H_N = - \frac{\hbar^2}{2 m } \: \sum_{ i = 1}^N \nabla^2_i + \sum_{ i  < j = 1}^N V ( r_{ij} ) + 
\sum_{ i = 1}^N U  ( {\bf r}_i ),
\label{nparh.1}
\eneq
\noindent
where ${\bf r}_i$ is the position of $i$th atom, $r_{ij}\equiv |{\bf r}_i-{\bf r}_j|$, and $V(r)$
 is the accepted (Aziz) interatomic pair potential for helium \cite{Aziz1979}.
The last term on the right-hand side of 
Eq.(\ref{nparh.1}) corresponds to the one-body potential describing the  the interactions of $^4$He atoms with the walls of a smooth cylindrical channel of radius $R$, representing the pore, whose axis is taken along the $z$ direction (specifically, it is the locus of all points with $x=y=0$). \\ \indent The expression for $U$ utilized here is described in detail in Ref. \onlinecite{Delmaestro2012}; it is derived from the well-known ``3-9" potential describing the interaction of a particle with an infinite, smooth wall, and it features an attractive well of depth $D$, as well as a repulsive core at short distance of characteristic length $a$.
These two parameters, together with the radius $R$, can be adjusted to reproduce, as closely as allowed by such a simplified model, the adsorption properties of real pores. In this work, we have made three different choices (summarized in Table \ref{table.1}), not  with the aim of reproducing any actual physical system, but rather of gaining understanding of the physics of $^4$He in rather different confining environments. The set ${\cal A}$ of potential parameters has been proposed to describe the interaction of helium atoms with a Na substrate \cite{Chizmeshya1998}, while ${\cal B}$ and ${\cal C}$ respectively have been used to model porous glasses  \cite{Delmaestro2012,Pricaupenko1995}. 
\\ \indent 
While the parameter sets labeled ${\cal B}$ and ${\cal C}$ may be more closely related to actual experimental systems (i.e., porous glasses), set ${\cal A}$ has been included with the aim of exploring the physics of the confined helium fluid in the presence of a weak adsorption potential. Alkali metal surfaces are known for their unusual adsorption properties \cite{Cheng1991,Boninsegni2004}; for example, a Cs substrate is not wetted at all by liquid helium, whereas a superfluid $^4$He monolayer has been predicted to form on a Li substrate \cite{Boninsegni1999}; Na is only slightly weaker an adsorber than Li. In confined geometries, alkali substrates have been show to enhance the superfluid response of nanoscale size fluid parahydrogen clusters \cite{Omiyinka2014}.

\begin{figure}
 \center
\includegraphics*[width=1. \linewidth]{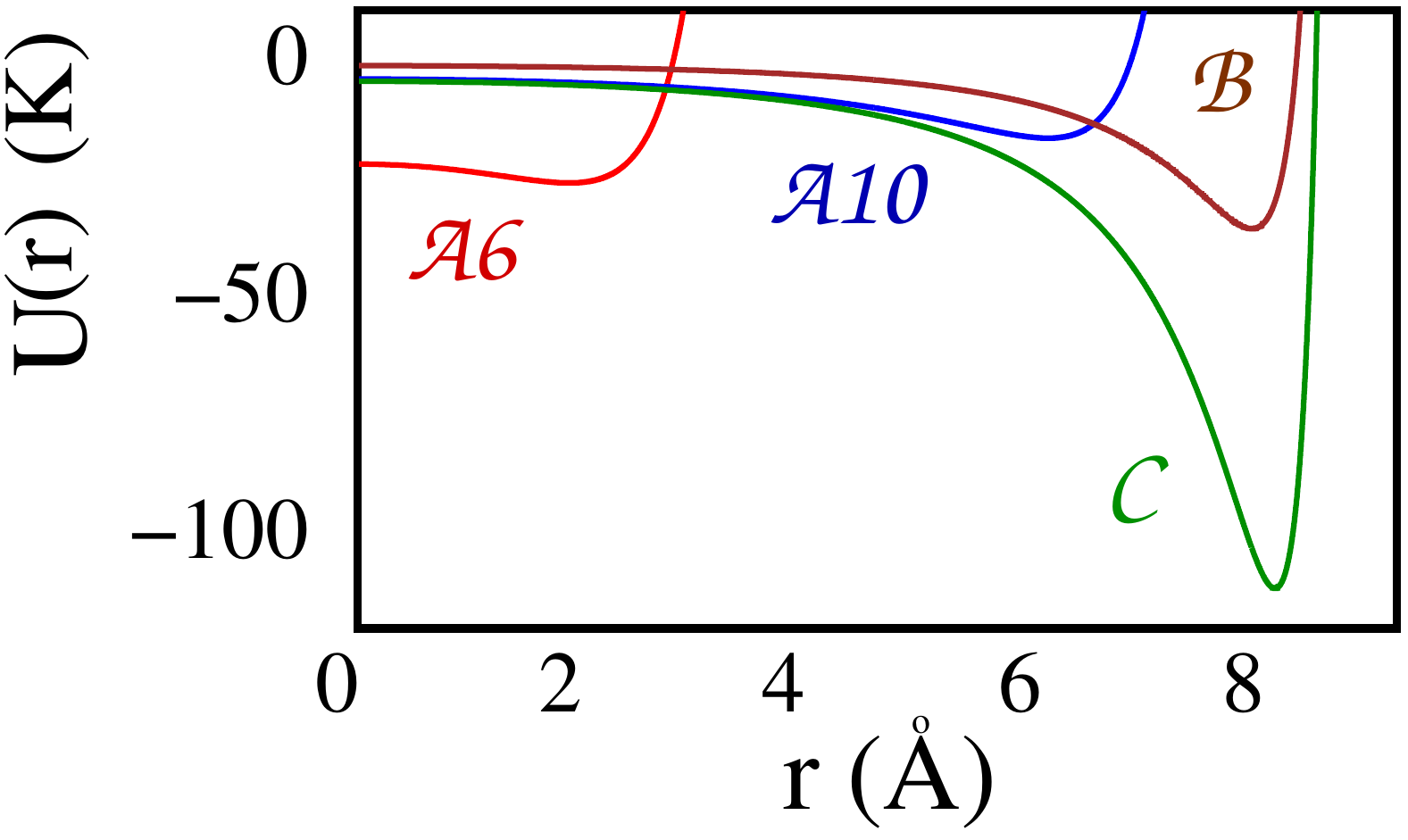}
\caption{Pore adsorption potential 
 corresponding to the parameters in Table \ref{table.1}, with pore radius $R$ set to 6 \AA\  ($\mathcal{A}6$) and 10 \AA\ 
($\mathcal{A}10$,\ $\mathcal{B}$ and  $\mathcal{C}$ ).
} 
\label{potentials}
\end{figure}

\begin{table}
\centering
\begin{tabular}{| c | c | c | }
\hline 
  {\rm System} & $D$ (K) & $a$ (\AA)  \\
\hline 
 $\mathcal{A}$ & 12.53 & 3.99 \\
\hline 
 $\mathcal{B}$  & 32 & 2.25  \\
\hline 
 $\mathcal{C}$ & 100 & 2.05  \\
\hline
\end{tabular}
\caption{Parameters $D$ (well depth) and $a$ (hard core range) of the $^4$He-wall adsorption  potentials ($U$ term in Eq. \ref{nparh.1}) adopted in this work.
The values for choice $\mathcal{A}$ (Na substrate) are taken from Ref. \onlinecite{Chizmeshya1998}, those for $\mathcal{B}$ (Glass 1) from Ref. \onlinecite{Delmaestro2012} and for $\mathcal{C}$ (Glass 2) from Ref. \onlinecite{Pricaupenko1995}. } 
   \label{table.1}
\end{table}
Fig. \ref{potentials} shows the various potentials considered in this work, for different pore radii. Clearly, substrate 
$\mathcal{C}$ is the strongest among those considered here, whereas, for a given radius $R$, the well depth of substrate $\mathcal{B}$ is roughly intermediate between that of the other two. The adsorption potential is nearly flat in the vicinity of the axis of the channel ($r=0$), but displays a minimum off the axis for $R\gtrsim 3$ \AA, whose depth decreases with increasing $R$. It is important to note that the potential labeled $\mathcal{A}6$ in Fig. \ref{potentials} is very similar to that used in Ref. \onlinecite{Markic2020}, the latter being shifted upward by about 23 K, compared to ours.
\\ \indent
The low temperature properties of the system described by (\ref{nparh.1}) were investigated in this work by means of QMC simulations based on the canonical \cite{Mezzacapo2006,Mezzacapo2007} continuous-space Worm Algorithm \cite{Boninsegni2006,Boninsegni2006_2}. As this methodology is extensively described in the original references, we shall not review it here. Details of the simulation are standard; the system in enclosed in a supercell shaped as a cuboid, with periodic boundary conditions in all directions \footnote{As the helium fluid is confined inside the channel, boundary conditions in the $x$ and $y$ are effectively irrelevant. The cell side in these directions is taken typically a few times $2R$.}. We made use of a fourth-order approximation for the high-temperature density matrix (see, for instance, Ref. \onlinecite {Boninsegni2005}), and all of the results quoted here are extrapolated to the limit of time step $\tau\to 0$. In general, we found that a value of the time step equal to $1.6\times 10^{-3}$ K$^{-1}$ yields estimates indistinguishable from the extrapolated ones. 
\\ \indent
The main physical quantities of interest are the radial $^4$He density profile $n(r)$, which provides information on the presence of one or more shells and their spatial definition, and the static structure factor $S(k)$, computed along the $z$ direction (i.e., the axis of the channel). The $S(k)$ is the key quantity that we use to assess the possible 1D nature of the fluid inside the pore, based on its scaling behavior with respect to $L$ and $T$, on the expectation that Haldane's model of Luttinger liquid ought to apply. %The  fundamental aspects of this formalism, which are directly relevant to the interpretation of our results, are outlined in Appendix \ref{harmonic_theory}.
\\ \indent
We also compute the superfluid fraction $\rho_S (L,T)$, estimated using the standard winding number estimator \cite{Pollock1987,Fisher1973}, with the goal of comparing our results to those of Ref. \onlinecite{Markic2020}, which are obtained using the same methodology \footnote{As superfluidity is underlain by quantum-mechanical exchanges of indistinguishable atoms taking place at low temperature, qualitative insight on the propensity of the system to flow without dissipation can also be obtained from the computed statistics of exchange cycles.}.
In principle, the scaling of $\rho_S (L,T)$ in the $L\to\infty, T\to 0$ limit, can also provide information about the 1D nature of the system; however, its calculation in the limit of long channels (i.e., the limit in which the most stringent  statements of the TLL apply)  quickly becomes a daunting task, due to the increasingly large fluctuations of the winding number and, correspondingly, the impractically long time required to accumulate the statistics needed to reduce error bars to a meaningful size.

\section{Results}
\label{4hepores}

In this section we present and discuss our 
results, chiefly focusing on the presence of TLL physics in $^4$He in nanopores, with the different geometries and adsorption potentials utilized.
\subsection{Dimensionality}
We begin by addressing immediately the central issue of this work, namely the possible crossover from 1D to 2D, and then from 2D to 3D behavior, which the authors of Ref. \onlinecite{Markic2020} claim take place in the pore as the radius and/or the density of helium inside are varied. In particular, based on a scaling analysis of the superfluid fraction $\rho_S 
(L,T)$, they assert that the crossover from 1D to 2D occurs as soon as the helium atoms inside the channel no longer form a single file along the axis, morphing into a cylindrical shell, even one of radius as small as $\sim$ 2 \AA.
\begin{figure}
 \center
\includegraphics*[width=1. \linewidth]{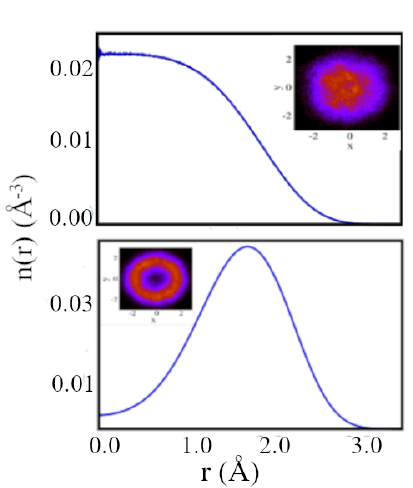}
\caption{{\it Top:} radial density profile $n(r)$, computed with respect to the axis of the pore. The linear $^4$He density is $n_0=0.25$ \AA$^{-1}$. Inset shows the $^4$He density projected on a plane along the axis of the cylinder (brighter colors mean higher density). This result is obtained with choice $\mathcal{A}$ of pore adsorption potential $U$, for a nominal pore radius $R=6$ \AA.\\
{\it Bottom:}  Same as top panel, but for linear density $n_0=0.6$ \AA$^{-1}$. }
\label{bulkd_25}
\end{figure}
\\ \indent 
Fig. \ref{bulkd_25} shows $^4$He radial density profiles computed at low temperature ($T=1$ K, they remain unchanged at lower temperature) for a helium fluid inside a pore of radius $R=6$ \AA, whose adsorption potential parameters are those of set $\mathcal{A}$ in Table \ref{table.1}, top (bottom) panel showing the result for  linear density $n_0=0.25 \ (0.60)$ \AA$^{-1}$. Also shown are projected density maps along the axis of the pore, illustrating, together with the $n(r)$, the structural change taking place as the linear density is increased, i.e., the helium atoms go from forming a single axial file at lower density, to a cylindrical shell of radius slightly less than 2 \AA\ at the higher $n_0$.
\\ \indent 
These results are {quantitatively very similar} to those of Ref. \onlinecite{Markic2020} for the same linear densities, albeit their results are generally obtained with different choices of pore radius and/or potential parameters. This similarity allows us to make a cogent  comparison of our results and predictions for these systems with theirs. 
For the purpose of our analysis, we momentarily postpone the discussion of the superfluid fraction, and present instead our results for the axial static structure factor $S(k)$.
 \begin{figure}
 \center
\includegraphics*[width=1. \linewidth]{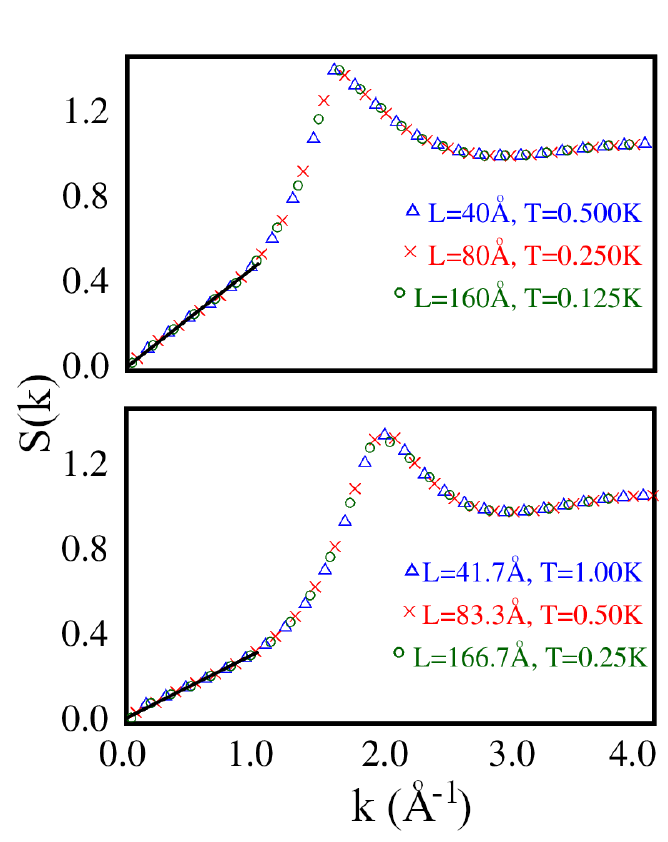}
\caption{ Static structure factor $S(k)$ computed along the axis of a pore of radius $R=6$ \AA, for a linear $^4$He pore density equal to $n_0=0.25$ \AA$^{-1}$ (top) and $n_0=0.6$ \AA$^{-1}$ (bottom). The parameters of the interaction between $^4$He atoms and the wall of the pore are those corresponding to set $\mathcal{A}$ in Table \ref{table.1}. Data shown pertain to three different system sizes; the temperature of each simulation is such that the product $LT$ is constant. Statistical errors are smaller than the sizes of the symbols. Solid line represents a fit of the data in the $k\to 0$ limit based on TLL theory, as explained in the text.}
 \label{sofk}
\end{figure}
\\ \indent
Fig. \ref{sofk} shows our QMC results for $S(k)$, for the two systems shown in Fig. \ref{bulkd_25}. The key result of TLL that we utilize to interpret our results, is the analytical expression for $S(k)$ 
%(Eq. \ref{corf.4} of Appendix \ref{harmonic_theory})
for a system of linear size $L$ at temperature $T$ \cite{Citro2007,Citro2008}. As mentioned in Sec. \ref{intro}, its main implications are that, in the $L\to\infty$, $T\to 0$ limits, $S(k)$ {\it a}) becomes a function of the product $LT$ alone {\it b}) behaves linearly as $k\to 0$.
\\
\indent
The results of Fig. \ref{sofk} clearly show that both these conditions are met by our numerical data. In particular, for each of the two values of $n_0$, the collapse of data obtained at constant $LT$ for three systems of different lengths (each length differing from the other two by at least a factor two), is downright impressive. Our  results are consistent with the linear behavior predicted at low $k$ by the TLL theory (solid lines in Fig. \ref{sofk}), allowing us to infer the value of the Luttinger parameter \cite{Kora2020}. 
%(Eq. \ref{corf.5}, see also Ref. \onlinecite{Kora2020}). 
We obtain $K\sim  
1.38$ (1.88) for $n_0=0.25$ ($n_0=0.60$) \AA$^{-1}$. These values of $K$, both in the $1 \le K\le 2$ range, are significantly lower than those reported, e.g., in Ref. \onlinecite {Delmaestro2010_b}; this is consistent with the weakness of the adsorption potential $\mathcal{A}$, compared to that utilized in those works (as it has already been shown for parahydrogen \cite{Omiyinka2016}, the strength of the potential $U$ affects the value of $K$).
\\
\indent 
The $S(k)$ shown in Fig. \ref{sofk} displays well-defined peaks in both cases. For the lower density system, i.e., $n_0=0.25$ \AA$^{-1}$, which as shown above is structurally quasi-1D, the peak is 
located at ${K}_1=2\pi n_0$, as expected. On the other hand, for the case $n_0=0.6$ \AA$^{-1}$, i.e., with $^4$He atoms arranged on a cylindrical shell, the location of the peak allows one to infer an {\it effective} 1D density roughly equal to $n_0/2$. One may interpret this by imagining the cylindrical shell as consisting of annuli coaxial with the pore, the distance between adjacent annuli being $2/n_0$, each annulus comprising on average two atoms, both located off the axis of the pore. 
\\ \indent
Altogether, the results of Fig. \ref{sofk} constitute strong evidence of 1D behavior in {\it both} cases, the only effect of the structural change undergone by the system as it evolves from a single file to a shell being the above-mentioned renormalization of the 1D density. There is no evidence whatsoever  of any dimensional crossover from 1D to 2D for the results obtained in this work at both linear densities are entirely consistent with TLL theory. Since, as stated above, we expect our results to be directly comparable to those of Ref. \onlinecite{Markic2020}, this apparent disagreement needs to be addressed.
 \begin{figure}[h]
 \center
\includegraphics*[width=1. \linewidth]{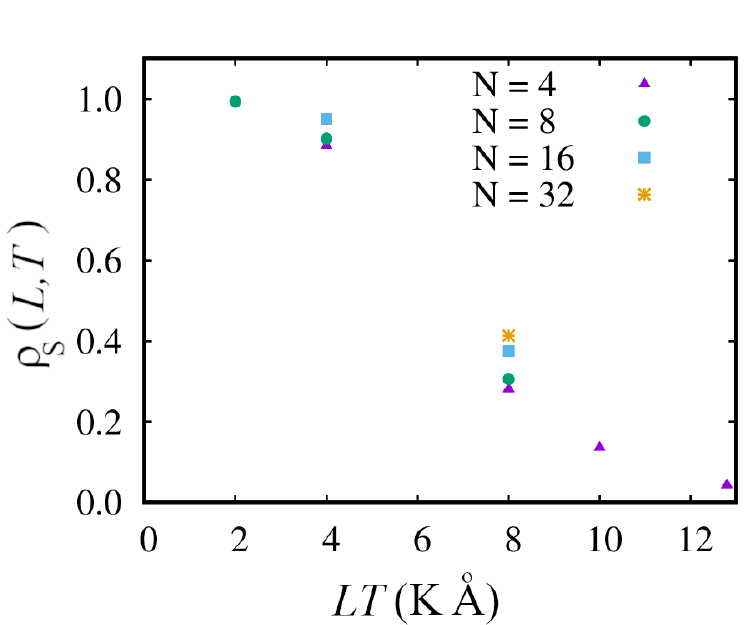}
\caption{ Superfluid fraction $\rho_S (L,T)$ for strictly  1D $^4$He at linear density $n_0=0.1$ \AA$^{-1}$, computed by 
QMC for different system sizes, plotted as a function of the product $LT$ of system size by temperature. $N$ is
the number of particles. Statistical errors on
 $\rho_S$ are of the order of 0.007, i.e., considerably smaller than symbol sizes. Finite-size effects are clearly seen.}
 \label{1dhe}
\end{figure}
\\ \indent
As  mentioned above, the contention of a dimensional crossover made in Ref. \onlinecite{Markic2020} is based on a scaling analysis of estimates of the superfluid fraction $\rho_S (L,T)$. We have computed $\rho_S 
(L,T)$ in this work as well, obtaining results which are in quantitative agreement with those of Ref. \onlinecite{Markic2020}, at least within the statistical errors of both calculations \footnote{The numerical agreement between our results for $\rho_S(L,T)$ and those of Ref. \onlinecite{Markic2020} is obviously scarcely surprising, as both calculations make use of the same computing methodology (in fact, the same computer code).}.
We argue that the findings of Ref. \onlinecite{Markic2020} are in fact consistent with TLL theory, if finite-size effects affecting estimates of $\rho_S(L,T)$ are properly taken into account.
It is well known that the superfluid fraction computed by QMC using the winding number estimator, is affected by finite-size effects, in {\it any} dimension; 1D is no exception. In 1D, finite-size corrections to the estimate of $\rho_S$ have been extensively studied in the context of the classical  XY
 model, which is a minimal model of superfluidity (see, for instance, Ref. \onlinecite{Hirashima2020}).
\\ \indent
Fig. \ref{1dhe} shows the superfluid fraction $\rho_S (L,T)$ computed for a strictly 1D system, namely $^4$He, at a linear density
 $n_0=0.1$ \AA$^{-1}$. The estimates are obtained in this work using the same QMC computational methodology, on systems comprising different numbers $N$ of particles, absolutely comparable to those utilized in Ref. \onlinecite{Markic2020}, and plotted as a function of the product $LT$. It is important to note that statistical errors on $\rho_S$ are of the order of 0.007, i.e., the obvious differences between results obtained for different numbers $N$ of particles are {\it well outside} statistical uncertainties. The trend is the same observed in Ref. \onlinecite{Markic2020}, i.e., for a given value of $LT$, estimates of the superfluid fraction obtained with a greater number of particles are greater in value. Thus, the results of Ref. \onlinecite{Markic2020}  {\it are entirely consistent with the} 1D {\it scenario}.
\\ \indent
The contention is made in Ref. \onlinecite{Markic2020} that an accurate fit to the estimates of the superfluid fraction obtained for the system of linear density $n_0=0.6$ \AA$^{-1}$ can be obtained by assuming a 2D scenario, i.e., in terms of an actual (BKT) superfluid transition occurring at a finite temperature (around 0.2 K). Besides the fact that, as noticed above, there is no reason to discard the 1D scenario, the interpretation of the QMC data in terms of a 2D BKT transition offered in Ref. \onlinecite{Markic2020} (Fig. 10) is unconvincing, as evidenced by the large finite-size effects affecting their estimates of $\rho_S(L,T)$ {\it below} the estimated transition temperature $T_c$, which are not at all typical of a BKT transition (see, for instance, Ref.  \onlinecite{Nguyen2021}).
\begin{figure}
\center
\includegraphics*[width=1. \linewidth]{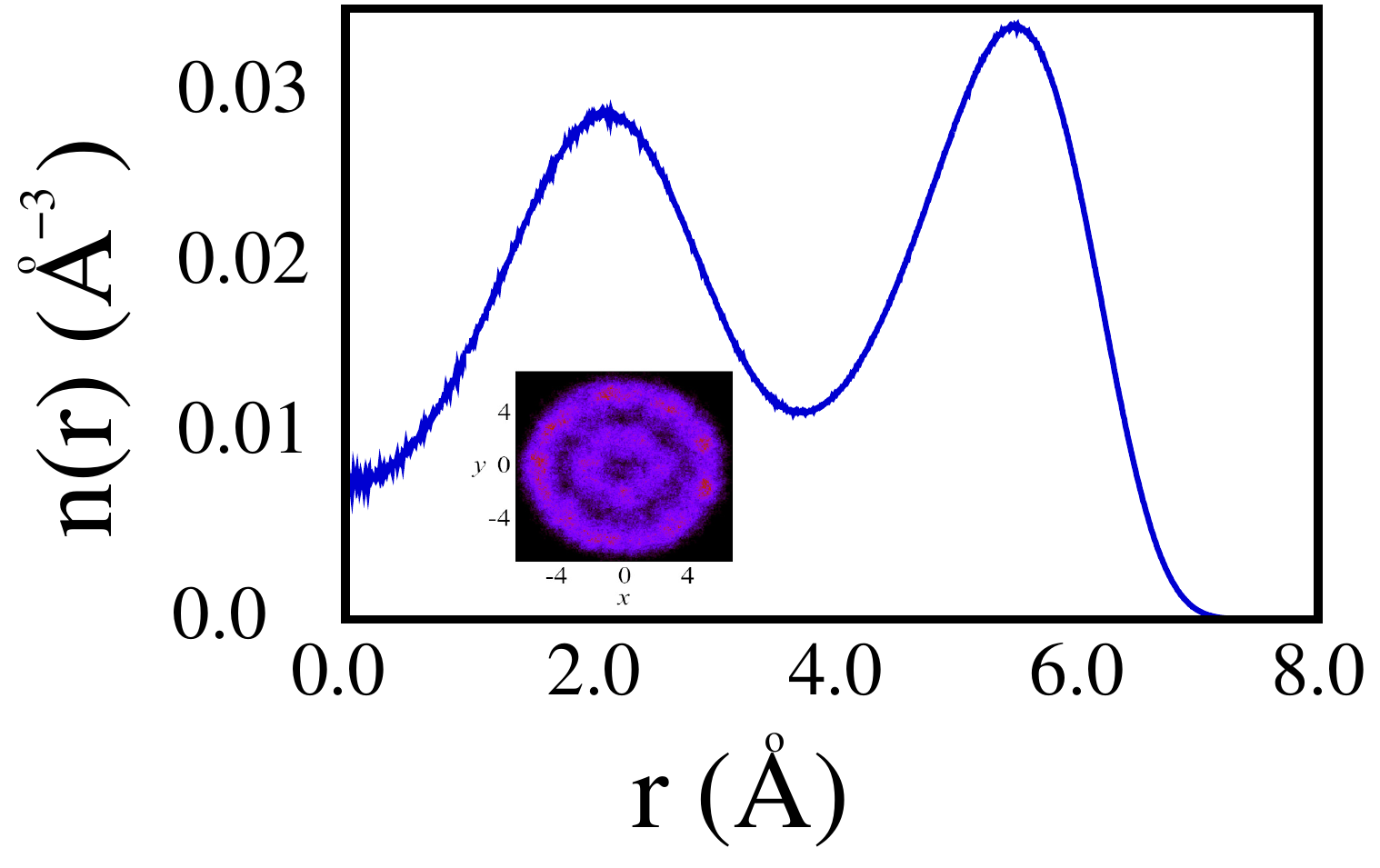}
\includegraphics*[width=1. \linewidth]{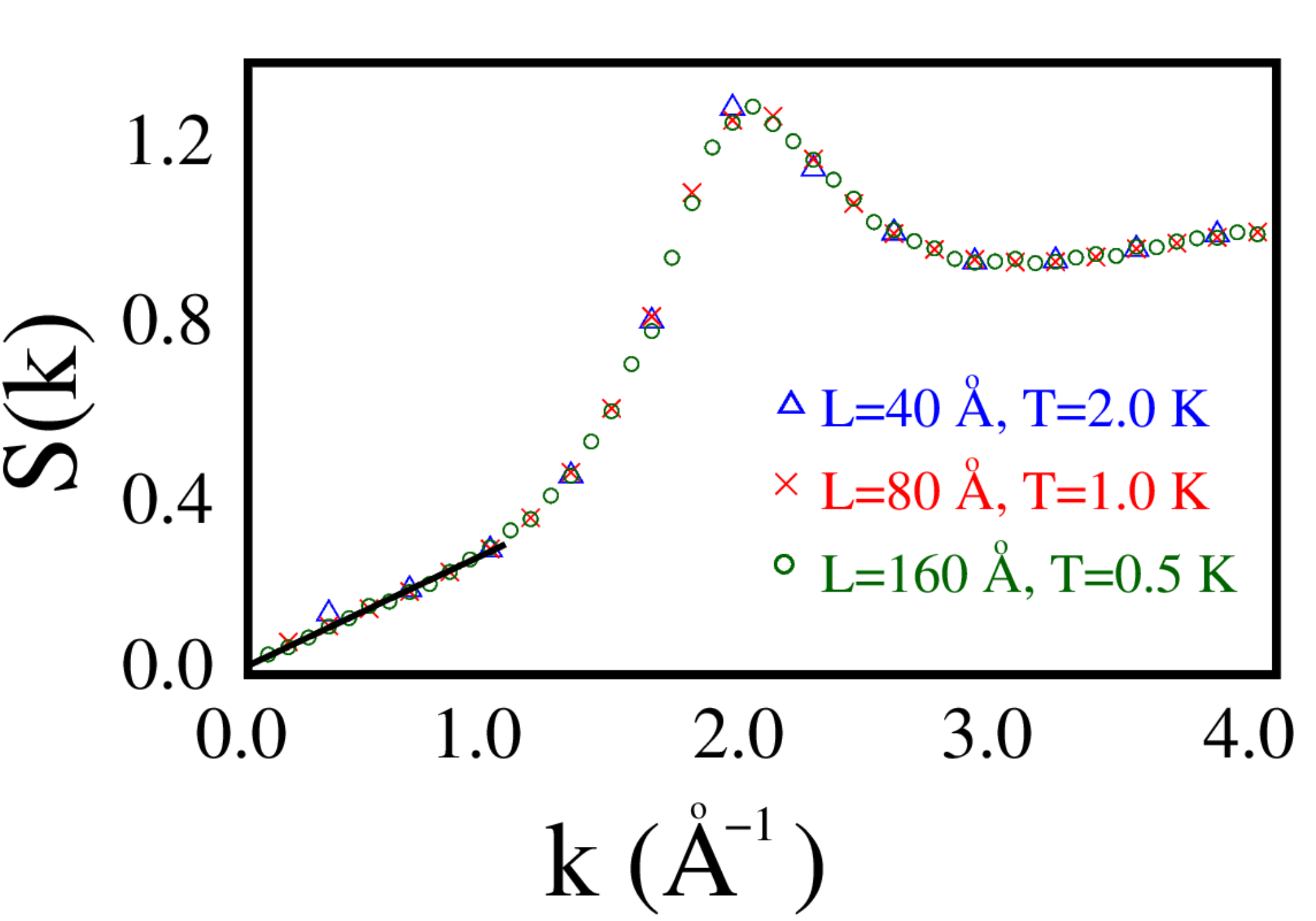}
\caption{{\it Top}: Radial density profile of a  $^4$He fluid of linear density $n_0=3$ \AA$^{-1}$, confined inside a pore of radius $R=10$ \AA. The result shown is at $T=1$ K, but does not change appreciably at lower $T$.
The interaction between $^4$He atoms and the wall of the pore are those of set $\mathcal{A}$ in Table \ref{table.1}. Inset shows the $^4$He density projected over a plane perpendicular to the axis of the cylindrical pore; brighter regions correspond to higher (3D) density.
{\it Bottom}: Static structure factor $S(k)$, computed along the pore axis for the same system, at three different system lengths and temperatures, such that $LT$ is held constant. Solid line is a linear fit to the data in the $k\to0$ limit, based on TLL theory.} 
\label{radial_3}  
\end{figure}
\\ \indent
One might wonder if a dimensional crossover may still take place, perhaps at different $n_0$ and/or pore radius. As we show below, the results obtained in this work, within the range of pore radii and density considered, are {\it all} amenable to a description in terms of the TLL theory, as $L\to \infty$. The crucial point, as we show below, is that one needs to perform calculations on systems of sufficient length, in order for the 1D physics to emerge. 
\\ \indent
Fig. \ref{radial_3} (top) shows the radial density profile $n(r)$ at $T=1$ K (it is independent of $L$ and remains unchanged at lower $T$, within the precision of the calculation) for a $^4$He fluid of linear density $n_0=3$ \AA$^{-1}$, confined inside a pore of radius $R=10$ \AA. The parameters of the adsorption potential are those of set $\mathcal{A}$ in Table \ref{table.1}. Clearly, in this case the fluid is nearly structure-less, with two floppy, largely overlapping concentric shells, arising essentially as a result of the hard core repulsion of $^4$He atoms at short distance. At low $T$ (i.e., $T \lesssim 2$ K) quantum-mechanical exchanges of helium atoms become frequent. An analysis of the number of particles involved in cycles of exchanges shows that atoms are delocalized between the two shells (or, as expressed in the formalism of Ref. \onlinecite{Yang2020}, ``hopping'' of helium atoms from one shell to the other occurs).
\\ \indent
It might be imagined that such a system ought not conform to the 1D paradigm, on account of the significant atomic motion in the transverse direction, especially since it paves the way to quantum-mechanical exchanges, which underlie SF in 2D and 3D $^4$He, but are increasingly suppressed as the 1D limit is approached. However, as shown in the bottom panel of Fig. \ref{radial_3}, our QMC-computed axial static structure factor displays the same data collapse described above for the single-file and single-shell systems, i.e., it
only depends on the product $LT$ for sufficiently long $L$, a hallmark of TLL behavior. We therefore conclude that even  this double-shell confined  $^4$He fluid behaves like a 1D system in the thermodynamic limit, one to which the formalism of Yang and Affleck \cite{Yang2020} ought to be applicable. The value of the Luttinger parameter $K$, in this case is $\sim 1.85$, again inferred from the low-$k$ behavior of $S(k)$, 
\\ \indent
It need be stressed that observing numerically the 1D limiting behavior of cylindrically-shaped 3D systems with a significant spatial extension in the radial direction,  requires that computer simulations be carried out on systems of sufficient length $L$. 
To give a sense of how crucial this aspect is, we discuss in detail the values of $\rho_S (L,T)$ at $T=1$ K, computed by QMC for the system of Fig. \ref{radial_3}, for the three quoted  lengths. For $L=40,\ 80$ \AA, i.e., with $N=120$ and 240 $^4$He atoms, the estimate of $\rho_S$ is the same within statistical uncertainties, namely 0.95(3). 
A value so close to unity, obtained at a relatively high temperature and apparently insensitive on $L$ might lead one to conclude that the system is behaving essentially like bulk $^4$He, i.e., a 3D SF. However, as soon as $L$ is increased to $160$ \AA, corresponding to $N=480$, 
$\rho_S$ falls to 0.70(3), showing how the estimates obtained with the shorter sizes do not offer a reliable representation of the behavior of the system in the $L\to\infty$ limit.
\\ \indent
It has been explicitly shown in the context of the 2D XY model  how the length $L$ of a quasi-1D system must be significantly greater than a characteristic length $L_c$, which grows proportionally to the perimeter (area) of an empty (filled) cylindrical shell, in order for the 1D behavior to emerge \cite{Yamashita2009,Hirashima2020}.
To illustrate this point, Fig. \ref{xy_all} shows numerical results for the superfluid density $\rho_S(L,T)$ of an XY ladder system defined on a $L \times M$  lattice, with periodic boundary conditions in both directions. As shown in the figure, the results obtained for different numbers $M$ of rungs collapse on the $M=1$ curve (i.e., a 1D system) if the length $L$ of the ladder is rescaled by $M$, $\rho_S  ( L , M , T )= \rho_S  ( L_{eff} , T )$, in the $L\to\infty$ limit.
\begin{figure}
\center
\includegraphics*[width=1 \linewidth]{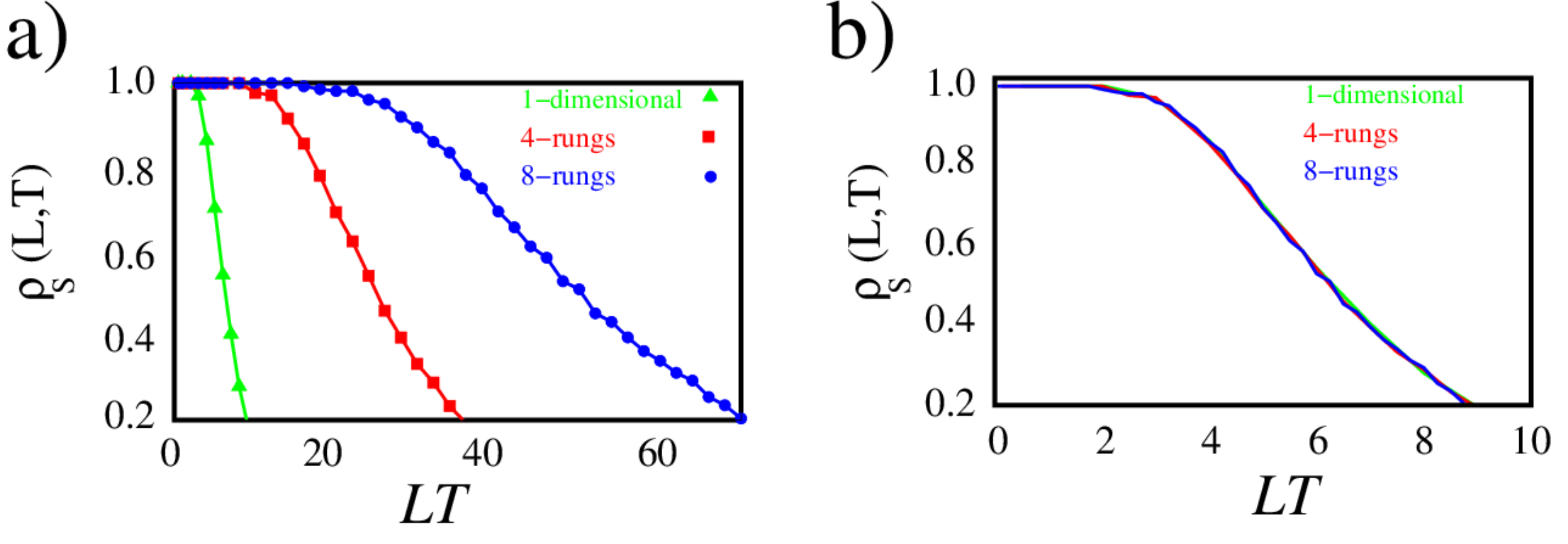}
\caption{Superfluid fraction $\rho_S  ( L ,  T )$ for a 2D XY ladder of length $L=4096$ and with $M$ rungs. Results are shown as a function of $LT$. Left panel shows results for 
$M=1$ (green curve - one-dimensional case), $M=4$ (red curve - 4 rungs), and $M=8$ (blue curve - 8 rungs). The same results are shown in the right panel, with the system length rescaled by  the corresponding $M$, showing the collapse of the curves onto each other.} \label{xy_all}  
\end{figure}
\noindent
%{\color{red} 
%Indeed, the XY model defined on a 2D $L \times M$  lattice, with periodic boundary conditions in both directions,  On recognizing that $M=1$ corresponds to the 1D case, a ladder system will approach the 1D limit as $L\gg M$ (i.e., $L_{eff}\gg 1$),  i.e., the scaling predictions for the 1D case will apply }. 
%The larger is $M$, the higher must be the value of $L$ in order for the finite-size effect to be negligible and for the system to crossover to the ``truly'' one-dimensional scaling regime. For finite $L$, the superfluid fraction may appear to display a quasi-2D behavior, but this is misleading as it reverts to 1D as soon as $L/M$ is made large enough \cite{Hirashima2020,Yang2020}.
%In order to obtain a numerical check of this result,  we computed  ${\rho_S  ( L , M,  T )} $ in an XY ladder for $M=1,4,8$. In Fig. \ref{xy_all}, we draw the unscaled and scaled plots for $L=4096$. The rescaled curves are practically indistinguishable from each other and the 1D behavior is observed.} 

\subsection{Role of adsorption potential}
\begin{figure}
    \centering
    \includegraphics{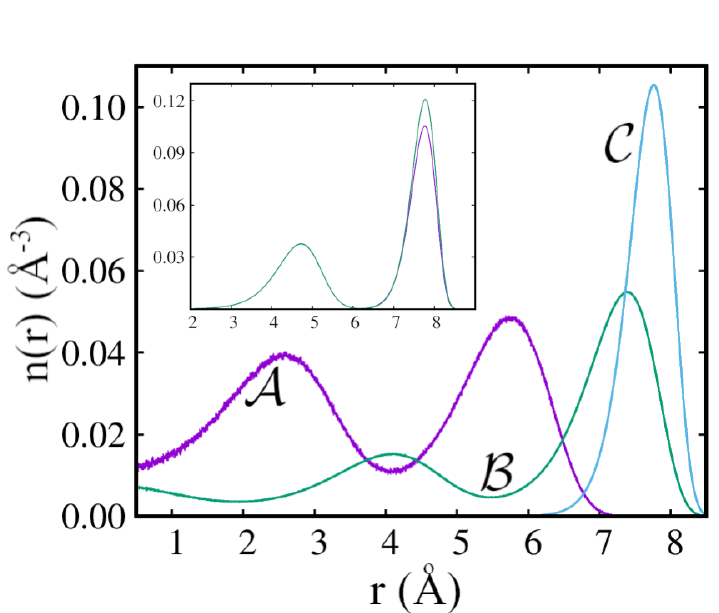}
    \caption{Radial density profiles for a $^4$He fluid in a cylindrical pore of radius $R=10$ \AA. for the three different adsorption potentials of Table \ref{table.1}, labeled accordingly. The linear $^4$He density in all cases is 4.0 \AA$^{-1}$. Inset compares profiles for case $\mathcal{C}$ of Table \ref{table.1} for linear densities 4.0 and 6.0 \AA$^{-1}$, the latter corresponding to the presence of two shells. All the results show are at temperature $T=1$ K, and remain unchanged at lower $T$.}
    \label{rcmp}
\end{figure}

We conclude this section by briefly discussing the dependence of the physics of the system on the adsorption potential. Generally speaking, if the radius of the pore is small (how small depends on the adsorption potential, but typically $R\lesssim 4$ \AA) only a single file of atoms forms
 along the axis; in the opposite limit, i.e., as $R$ becomes large, adsorption of $^4$He inside the pore approaches that on a flat substrate, i.e., continuous growth of a superfluid $^4$He film as a function of chemical potential, on top of possibly one or few ``inert'', solid layers, depending on the strength of the substrate.
Inside a pore of diameter of the order of few nm, interesting, intermediate behavior can be observed, as adsorption
begins to take place near the surface, and proceeds through the formation of concentric shells. The degree of floppiness of each shell, and the corresponding overlap of adjacent shells (and quantum-mechanical exchanges of atoms at low $T$)  depend on both the value of $R$ and on the adsorption potential.
\\ \indent
Fig. \ref{rcmp} shows QMC-computed radial density profiles of $^4$He inside a cylindrical pore of radius $R=10$ \AA, for the  three different adsorption potentials whose parameters are summarized in Table \ref{table.1}. The $^4$He linear density $n_0 = 4$ \AA$^{-1}$. As one can see, as the depth of the attractive well of the potential is increased, and concurrently the distance of closest approach $a$ decreases, the outer shell becomes sharper, forms closer to the substrate, and can pack a greater number of $^4$He atoms.\\ \indent  For the particular case of potential $\mathcal{C}$, which is the most strongly adsorbing of the three, only one shell forms, coating the surface of the pore, of effective 2D density over 0.08 \AA$^{-2}$, i.e., above the 2D freezing density of $^4$He \cite{Gordillo1998}. Atoms confined within this shell experience very little mobility and exchanges are virtually non-existent. On increasing the $^4$He density inside the pore, 
{ a second shell forms, well separated from one another and with no atomic exchanges (inset of Fig. \ref{rcmp}). On further increasing the density a central file of atoms may or not appear,  depending on  geometry and adsorption potentials, both affecting commensuration.}
\\ \indent
Very different behavior is observed on the weaker $\mathcal{A}$ and $\mathcal{B}$ substrates, in which the outer shell that coats the surface forms further away from it, and the shells that form are broad, floppy and liquid-like, and overlap. For a given choice of radius and for a specific linear density, the value of the Luttinger parameter $K$ is lower for weaker adsorption potentials.

\section{Conclusions}
\label{conclusions}
We have carried out in this work extensive QMC simulations at low temperature (typically 1 K or less) of a fluid of $^4$He atoms in the confines of a cylindrical pore of axial length greatly exceeding its radius. We have considered different values of the radius, up to 1 nm, and three different adsorption potentials, with greatly varying adsorption strength. The main purpose of this study was to assess recent contentions of dimensional crossover(s) taking place inside cylindrical pores, including for values of the pore radius as small as 4 \AA, in correspondence of the formation of a single shell (as opposed to a single file of atoms along the axis).\\
\indent
Our analysis of the static structure factor, computed along the axis of the pore, shows data collapse in the limit of pore length $L\to \infty$, consistently with TLL theory, i.e., the accepted 1D paradigm, for all values of the linear $^4$He density $n_0$, pore radius $R$ and adsorption potentials utilized in this work. We also find this to be true regardless of the structure of the fluid inside the pore, i.e., whether it is a single file of atoms along the axis, a single hollow shell, or a fluid filling a section of the pore, in which few overlapping shells can be identified. This is consistent with the formalism of Yang and Affleck, which models the multi-shell structure of the confined fluid as a system of coupled  TLLs, still exhibiting overall 1D behavior \cite{Yang2020}. 
\\ \indent
We find the value of the Luttinger parameter $K$ to fall in the range between 1 and 2 for the two weaker adsorption potentials considered, i.e., considerably less than what found in previous work, in which stronger adsorption potentials were used. This is consistent with what found in other calculations \cite{Omiyinka2016}, namely that adsorption in confined geometries characterized by weak potentials can reduce the value of $K$, i.e., bring a fluid close to a topologically protected, quasi-superfluid phase, characterized by $K < 0.5$. Whether a further, substantial reduction of $K$ could be achieved by selecting even weaker substrates, e.g., Cs \cite{Chizmeshya1998} will be the subject of future studies.
\\ \indent
We find that, as long as $L/R \gg 1$, the system always displays 1D behavior, i.e., we find no dimensional crossover.
We attribute the disagreement between our conclusions and those of Ref. \onlinecite{Markic2020} to the smallness of the system length considered, and to the incorrect assumption
that numerical data for the superfluid fraction should not be affected by finite-size corrections. In our submission, the data presented in Ref. \onlinecite{Markic2020} are consistent with 1D behavior.
\\ \indent
As a final remark, the possibility of clearly identifying TLL behavior in $^4$He in nanopores paves the 
way to a plethora of possible extension of our research, toward, for instance, constructing pertinently
designed junctions and/or networks of pores, where to realize in a tunable and controlled system 
the novel phases and phase transitions predicted in junctions of TLLs \cite{Chamon2003,Oshikawa2006,giuliano_dual,Hou2012,Giuliano2007,Giuliano2019,Kane2020}, 
including the ones involving  fixed points at reduced decoherence \cite{Novais2005,Giuliano2008}, with potentially countless applications. 
 
 \vspace{0.2cm}
 
{\bf Acknowledgements:} We thank Ian Affleck and Arturo Tagliacozzo for a critical review of the manuscript. This work was supported in part by the Natural Sciences and Engineering Research Council of Canada (NSERC). M. B. gratefully acknowledges the hospitality of the Universit\`a della Calabria, where most of the research work was performed. Computing support of ComputeCanada is acknowledged as well. A. N. was financially supported  by POR Calabria FESR-FSE 2014/2020 - Linea B) Azione 10.5.12,  
grant no.~A.5.1.  D.G.   acknowledges  financial support  from Italy's MIUR  PRIN project  TOP-SPIN (Grant No. PRIN 20177SL7HC).

\normalem
\bibliography{helium.bib}
\end{document}